\documentclass[prl,twocolumn,superscriptaddress,showpacs]{revtex4}% Physical Review Letters
\usepackage{graphicx,amsmath,epsf}

\newcommand{\vv}{\mathbf{v}}
\newcommand{\vR}{\mathbf{R}}

\begin{document}
\title{Ehrenfest-time dependence of weak localization in open quantum dots}

\author{Saar Rahav}
\affiliation{Laboratory of Atomic and Solid State Physics, Cornell University, Ithaca 14853, USA.}
\author{Piet W. Brouwer}
\affiliation{Laboratory of Atomic and Solid State Physics, Cornell University, Ithaca 14853, USA.}

\begin{abstract}
Semiclassical theory predicts that the 
weak localization correction to the conductance of a ballistic
chaotic cavity is suppressed if the Ehrenfest time
exceeds the dwell time in the cavity [I.\ L.\ Aleiner and A.\ I.\
Larkin, Phys.\ Rev.\ B {\bf 54}, 14424 (1996)]. 
We report numerical simulations
of weak localization in the open quantum kicked rotator that confirm
this prediction. Our results disagree with the `effective random
matrix theory' of transport through ballistic chaotic cavities.
\end{abstract}
% other possible PACS number: 73.63.Kv Quantum dots
\maketitle

The wave nature of electrons is the cause of striking effects that are
absent in classical mechanics. The best known manifestations of quantum
mechanics on electrical transport are weak localization (WL), universal
conductance fluctuations (UCF), and shot
noise~\cite{kn:beenakker1991b}. In this letter we consider such
quantum mechanical effects in a ballistic cavity with chaotic
classical dynamics. 
%In that case, the signatures of quantum transport 
%become universal: they are independent of details of the motion inside 
%the cavity. 
A phenomenological theory of shot noise, weak
localization, and
conductance fluctuations in a ballistic chaotic cavity is given 
by random matrix theory \cite{kn:beenakker1997}. 
The predictions of random matrix theory have been confirmed using
semiclassical methods 
\cite{kn:argaman1995,kn:aleiner1996,kn:takane1998,kn:blanter2000}. 

Since they are intrinsically quantum effects, it is legitimate to ask
what happens to shot noise, weak localization, and conductance
fluctuations if the classical limit is taken. The classical limit 
corresponds to the case that the electron wavelength is much smaller
than the system size. The relevant time scale for this question is
the so-called ``Ehrenfest time'' \cite{kn:zaslavsky1981,kn:aleiner1996}, 
the minimal time at which the wave 
nature of electrons become apparent. This is the time at which a 
minimal wavepacket is stretched to the system size $L$ (or the size 
of the lead opening). It is given by
\begin{equation}
  \tau_{\rm E} = \lambda^{-1} \ln k L,
\end{equation}
where $\lambda$ is the Lyapunov exponent characterizing the chaotic
motion in the cavity, and $k$ the electron wavenumber.

The effects of the Ehrenfest time are most prominent if $\tau_{\rm E}$
is larger than the mean dwell time in the cavity $\tau_{\rm D}$. In 
this regime most of the transport is classical
and shot noise is 
suppressed~\cite{kn:beenakker1991c,kn:agam2000,kn:oberholzer2002}. 
The effect of $\tau_{\rm E}$ on WL was first 
addressed by Aleiner and Larkin~\cite{kn:aleiner1996}, who
predicted a suppression proportional to $\exp(-2 \tau_{\rm
  E}/\tau_{\rm D})$. Exponential suppression of WL, but
with a different 
exponent, was also found by Adagideli~\cite{kn:adagideli2003}.
Suppression of WL was observed experimentally by Yevtushenko 
{\em et al.} \cite{kn:yevtushenko2000}.
There is no 
semiclassical theory of the Ehrenfest-time dependence of UCF.

Numerical simulation of systems with large Ehrenfest times
is difficult due to the need for (exponentially)
high values of $kL$. To reduce numerical costs Jacquod {\em et al.}
proposed to replace the two-dimensional cavity by a 
one-dimensional quantum map~~\cite{kn:jacquod2003}.
The map is `opened', so that
simulation of transport properties is possible. An example
of such a map, the open kicked rotator, was used to obtain
numerical results for shot noise~\cite{kn:tworzydlo2003},
WL~\cite{kn:tworzydlo2004b}, and UCF
~\cite{kn:jacquod2004,kn:tworzydlo2004,kn:tworzydlo2004c}.
The simulation results for shot noise were in agreement
with semiclassical predictions~\cite{kn:agam2000}. In contrast, the results for
WL and UCF are remarkably different: no dependence 
on $\tau_{\rm E}$ was found.
To explain their results, the authors of Refs.\
\cite{kn:tworzydlo2004b,kn:jacquod2004,kn:tworzydlo2004,kn:tworzydlo2004c}
proposed an `effective random matrix theory': Noting that
quantum diffraction contributes only if the classical
dwell time is larger than $\tau_{\rm E}$, they proposed to apply
random matrix theory only to the part of phase space with dwell
times longer than $\tau_{\rm E}$ \cite{kn:silvestrov2003}. Since
UCF and WL are not dependent on the size of the phase space 
involved, the `effective random matrix' theory is able to 
describe the numerical results for WL and UCF \cite{foot1}.
%This phenomenological description is in contrast with
%semiclassical calculations of WL. 

\begin{figure}
\epsfxsize=0.7\hsize
\epsffile{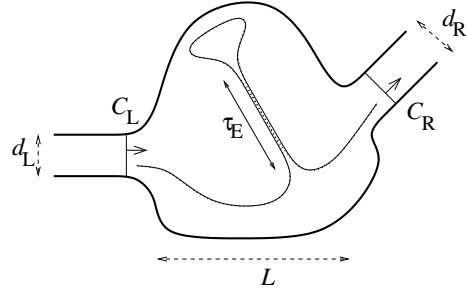}

\caption{\label{fig:1} 
Schematic picture of a ballistic chaotic cavity of size
$L$. The cavity is attached to two leads, labeled ``L'' and ``R'',
and with width $d_{\rm L}$ and $d_{\rm R}$, respectively. The 
contours $C_{\rm L}$ and $C_{\rm R}$ drawn in the 
leads are used for the calculation of conductance. A schematic 
drawing of relevant trajectories for weak localization is depicted
inside the cavity. (The physical trajectories scatter from the
cavity walls many times.) }
\end{figure}

Apart from the prediction of the {\em size} of the weak localization
correction, the `effective random matrix theory' and the
semiclassical theory have different predictions of the minimal time 
required for weak localization to occur. In the semiclassical theory, 
quantum interference requires 
a minimal wavepacket to be split {\em and} reunited,
which takes a minimal time $2 \tau_{\rm E}$. A schematic diagram
drawing relevant semiclassical trajectories for WL is depicted in 
Fig.\ \ref{fig:1}. This is different in
the 'effective random matrix theory, where quantum interference
is fully established already after a time $\tau_{\rm E}$. 

It is the purpose of this letter
to resolve the discrepancy between the reported numerical simulations
and the semiclassical theory for weak localization. In the first
part of this letter, we review the original 
semiclassical theory \cite{kn:aleiner1996}
of the Ehrenfest time dependence of WL and amend it in order to
correctly implement
classical correlations in an open cavity
\cite{foot2}. 
In the second half we report 
time-resolved numerical simulations of weak localization in the
open quantum kicked rotator. The time-resolved simulations allow
us to determine the time at which WL first appears, as well as the
effect of a finite Ehrenfest time on the conductance. We find
that the time-resolved simulations for WL and the corrected semiclassical
theory are consistent.

{\em Semiclassical theory.} The semiclassical theory of 
Ref.\ \onlinecite{kn:aleiner1996} considers a two-dimensional chaotic 
cavity which is coupled to two electron reservoirs via contacts
of widths $d_{\rm L,\rm R}$, see Fig.\ \ref{fig:1}. One has
$d_{\rm L,\rm R} \ll L$, so that electron dynamics inside the
cavity is ergodic.
The system is studied in the semiclassical limit, where the number of
open channels is large, $d_{\rm L,\rm R}k \gg 1$. This separation of
length scales is equivalent to the following separation of time scales
$(v k)^{-1} \ll L/v \sim 1/\lambda \ll \tau_{\rm
  D}$, where $v$ is the electron velocity, and
the dwell time is $\tau_{\rm D} = \pi {\cal A}/(d_{\rm L} + d_{\rm R})
v$, ${\cal A}$ being the area of the cavity.
It will be useful to define the probability that an electron at a random
point will leave the cavity through contact $i$, 
$P_i=d_i/(d_{\rm R}+d_{\rm L})$.

The central objects in the semiclassical theory of Aleiner and Larkin 
are the non-oscillating
parts of the product of advanced and retarded Green functions
\cite{kn:aleiner1996}.
The ``diffuson'' ${\cal D}$ is composed of combinations of orbits with
themselves while the ``cooperon'' ${\cal C}$ is composed of orbits and 
their time reversed counterparts.
Quantum effects are introduced by including a weak additional 
random quantum potential.
This random potential leads to two different effects: 
phase space diffusion, and coupling between paths which are close in phase
space. Both are expected due to quantum uncertainty. After averaging
over the random potential, the equation of motion for the 
leading order diffuson, ${\cal D}^0$, is~\cite{kn:aleiner1996}
\begin{equation}
\label{diffusonw}
\left[ -i \omega + \hat{L}_1 - \gamma_{\rm q} \frac{\partial^2}{\partial \phi_1^2} \right] {\cal D}^0 ( \omega ; 1 ,2 ) = \delta(1,2),
\end{equation}
where $j\equiv (\phi_j, {\bf R}_j)=1,2$ denotes the phase space coordinates,
limited to the energy shell: ${\bf R}$ is the electron 
position while the angle $\phi$ is the direction of its
velocity $\vv$. 
The strength of the phase space diffusion is $\gamma_{\rm q} \simeq \lambda^2/kv$.
The operator $\hat{L}_1$ is the Liouville operator.
For a hard wall cavity, $\hat{L}_1= \vv_1 \cdot {\partial}/{\partial
  \vR}_1$, with appropriate boundary conditions at the walls.
The symbol $\delta(1,2)\equiv 2 \pi
\delta ( {\bf R}_1-{\bf R}_2) \delta ( \phi_1-\phi_2)$ denotes a delta function on the energy shell. 
The leading order cooperon, ${\cal C}^0$, also satisfies
Eq.\ (\ref{diffusonw}).

The WL correction to the diffuson arises from combinations
of a trajectory with small angle self-intersection and a
partner without such self-intersection~\cite{kn:aleiner1996,kn:sieber2001},
see Fig.~\ref{fig:1}. The 'interference region' is composed
of close phase space points (with distance $\sim k^{-1/2}$) where 
quantum diffraction is important. This leads to a quantum correction
for the diffuson ${\cal D} = {\cal D}^0 + \Delta {\cal D}$ 
with~\cite{kn:aleiner1996}
\begin{multline}
\label{dd}
\Delta {\cal D} (1 , 2) =  {\cal D}^0 (1 , \bar{2}) 
\frac{{\cal
    C}^0 (\bar{2} , 2)}{2 \pi \hbar \nu} +  \frac{{\cal C}^0 (1,
  \bar{1})}{2 \pi \hbar \nu} {\cal D}^0 (\bar{1}, 2 ) 
 \\ -  \int d3 {\cal D}^0 (1,3) {\cal D}^0 (\bar{3},2) 
 \left[ \hat{L}_3 - \gamma_{\rm q} \frac{\partial^2}{\partial \phi_3^2} \right] \frac{{\cal C}^0 (3,\bar{3})}{2 \pi \hbar \nu},
\end{multline}
where $\bar{j} = (\phi_{j}+\pi, {\bf R}_j)$ denotes the time reversal of 
phase space point $j$ and 
$\nu={m}/{2 \pi \hbar^2}$ is the density of states per unit area.
Since we consider DC transport, the frequency $\omega$ has been
set to zero.

The total (DC) conductance of the dot can be expressed in terms of 
the diffuson 
${\cal D}$~\cite{kn:aleiner1996,kn:takane1998},
\begin{multline}
\label{totT}
G = 2 \pi \hbar \nu v^2 \int_{C_{\rm R}} dl_1 
  \int_{\phi_{\rm R}-\pi/2}^{\phi_{\rm R}+\pi/2} \frac{d \phi_1}{2 \pi}
  \cos(\phi_{\rm R} - \phi_1) \\ \mbox{} \times \int_{C_{\rm L}} dl_2
  \int_{\phi_{\rm L}-\pi/2}^{\phi_{\rm L}+\pi/2} \frac{d \phi_2}{2 \pi}
  \cos(\phi_{{\rm L}} - \phi_2)
  {\cal D} (1,2).
\end{multline}
Here $G$ is measured in units of the conductance quantum $2 e^2/h$,
and 
$C_{\rm L}$ and $C_{\rm R}$ denote cross section contours of the contacts.
 The angle $\phi_{\rm L}$ is
the direction of the inward-pointing normal 
to $C_{\rm L}$, while $\phi_{\rm R}$ is the outward-pointing normal to
 $C_{\rm R}$.

The WL correction to the conductance $\delta G$ 
is obtained by substituting Eq.\ (\ref{dd}) for $\Delta {\cal D}$ 
into Eq. (\ref{totT}).  The first two terms in Eq. (\ref{dd}) 
contain paths which leave the dot at the point of entry, and hence
do not contribute to the transmission. The phase space point 3 in
the third term in Eq.\ (\ref{dd}) can be viewed as the center of the 
'interference region' depicted in Fig.~\ref{fig:1}. As the two 
diffusons and the cooperon head into opposite directions in 
configuration space, one can treat them as statistically uncorrelated
and estimate them independently~\cite{kn:aleiner1996}.

Both the cooperon and the product of diffusons in Eq.\ 
(\ref{dd})
exhibit non-trivial correlations \cite{kn:aleiner1996}, 
since the incoming 
path is strongly correlated with the outgoing one, cf.\ 
Fig.\ \ref{fig:1}.  
An intuitive way to estimate the cooperon (or the product of diffusons)
was presented by Vavilov and Larkin \cite{kn:vavilov2003}.
They demonstrated that the effect of phase space diffusion is
equivalent to an average over phase space regions of size $\sim k^{-1/2}$
near $3$ and $\bar{3}$. Following Ref.\ \onlinecite{kn:vavilov2003},
we consider phase space points $3'$ and $3''$ within a
distance of order $k^{-1/2}$ from $3$. 
The phase space distance between classical orbits starting at the points 
$3'$, $3''$ will diverge exponentially due to the chaotic dynamics and 
will be macroscopic after a time $\tau_{\rm E}/2$.
Once the phase
space distance between the trajectories is large, they can be 
considered uncorrelated. 

For the cooperon, the classical correlations mean that it takes
a minimal time to form a closed loop. Let us denote by $t_j$
the time it takes for a classical trajectory starting
at phase space point $j$ to leave the cavity. If $t_{\bar{3}}<\tau_{\rm E}/2$
all the points near $\bar{3}$ leave the cavity before it is possible
to close a loop. In contrast, when $t_{\bar{3}}>\tau_{\rm E}/2$,
the cooperon is estimated by propagating phase space points
close to $\bar{3}$ forward in time, and points close to $3$
backward in time, for a time of $\tau_{\rm E}/2$. After
this propagation the points which contribute to the local average
are uncorrelated. Then the cooperon can be estimated using its
ensemble average $\tau_{\rm D}/{\cal A}$. The resulting
cooperon in the open cavity is given by
\begin{equation}
\label{cooperon}
 {\cal C}^0 (3,\bar{3})= \left<{\cal C}^0 (3'',\bar{3'}) \right> \simeq \frac{\tau_{\rm D}}{\cal A} \theta(t_{\bar 3} - \tau_{\rm E}/2),
\end{equation}
where $\theta(x) = 1$ if $x > 0$ and $0$ otherwise. 

The product of diffusons is estimated similarly. The phase space
points 1 and 2 in Eq.\ (\ref{dd}) refer to two different contacts.
If $t_3 < \tau_{\rm E}/2$, trajectories starting close to $3$ are 
still correlated when 
they leave the cavity. Therefore, they cannot arrive at points
on {\em different} contacts, leading to a vanishing contribution
to $\delta G$. 
When $t_3 > \tau_{\rm E}/2$ the lead integrals will result in a 
factor of $P_R P_L$, as in Refs.\
\onlinecite{kn:aleiner1996,kn:takane1998}.
Note that the WL paths are composed of four 'legs' with minimal time
of $\tau_{\rm E}/2$, so that the minimal time of the whole path
is $2\tau_{\rm E}$.

The resulting WL correction is
\begin{equation}
  \delta G = - P_{\rm R} P_{\rm L} \frac{\tau_{\rm D}}{\cal A}
  \int d3\, \theta(t_3 - \tau_{\rm E}/2)  \hat L_3
  \theta(t_{\bar 3} - \tau_{\rm E}/2).
  \label{eq:deltat2}
\end{equation}
The Liouville operator in Eq.\ (\ref{eq:deltat2}) measures the rate of
flow of probability density out of the integration range of the
phase space variable $3$~\cite{kn:aleiner1996}. The
boundary of the integration range is composed of the lead phase space 
points propagated {\em backward} for a time $\tau_{\rm E}/2$. This leaves 
only a fraction $\exp({-\frac{\tau_{\rm E}}{2\tau_{\rm D}}})$ 
of the size of the
boundary at the lead. 
In addition, 
only points which also have $t_{\bar{3}}>\tau_{\rm E}/2$ will have a 
non-vanishing cooperon, leading to another
 factor of $\exp({-\frac{\tau_{\rm E}}{2\tau_{\rm D}}})$. Hence,
we find
\begin{equation}
\delta G = - P_{\rm R} P_{\rm L} e^{-\tau_{\rm E}/\tau_{\rm D}}.
  \label{eq:dtresult}
\end{equation}
The result (\ref{eq:dtresult}) shows that, even after accounting
for classical correlations that were not accounted for in Ref.\
\onlinecite{kn:aleiner1996}, the WL correction to transmission 
is suppressed exponentially. However, the exponent is
different from that of Ref.\ \onlinecite{kn:aleiner1996},
where it is reported that $\delta G \propto 
\exp(-2 \tau_{\rm E}/ \tau_{\rm D})$. 

%The WL correction for reflection
%can be computed in a similar manner, with the result $\delta R =
%-\delta T$. For details, we refer to Ref.\ \onlinecite{kn:rahav2006}.

{\em Numerical simulations.} We have compared the predictions of the 
semiclassical theory to numerical simulations of the open quantum
kicked rotator. The time evolution of the kicked rotator is discrete,
and given in terms of the Floquet operator $\psi (t+1) = {\cal F} \psi 
(t)$. The state of the system $\psi$ is a finite vector of dimension 
$M$. The Floquet operator is an $M \times M$ matrix and reads
\begin{eqnarray}
  F_{nm} &=& M^{-1/2} e^{-i \pi/4 + i \pi (m-n)^2/M}
  \\ && \mbox{} \times e^{ - i (\kappa M/4 \pi)
  (\cos(2 \pi m/M+\theta) + \cos(2 \pi n/M+\theta))}, \nonumber
\end{eqnarray}
where $\kappa$ is a numerical parameter that determines the classical
dynamics of the map and $0 \le \theta < 2 \pi/M$ a parameter that
sets the precise quantization condition. The region $\kappa 
\gtrsim 7.5$ is associated with
classically chaotic dynamics. 
The size of the matrix $M$ is taken to be even.

To open the system, two consecutive sets of $N$
elements of $\psi$ are chosen as the contacts \cite{kn:jacquod2003}. 
A scattering matrix
is then constructed by the rule
\begin{equation}
 S(\varepsilon) = \sum_{t=1}^\infty e^{i \varepsilon t} S(t);
 \ \ \
 S (t) = P \left[ {\cal F} Q \right]^{t-1} {\cal F} P^T,
  \label{eq:Smat}
\end{equation}
where $\varepsilon$ is the quasienergy, $P$ is a $2 N \times M$ matrix
projecting on the contact sites, and $Q \equiv 1 - P^T P$. 
The semiclassical limit
corresponds to $M \rightarrow \infty$ while keeping $\tau_{\rm
  D}=M/2N$ fixed. For large values of $\kappa$ 
the Lyapunov exponent $\lambda$ can be approximated $\lambda \simeq 
\ln \left( \kappa/2 \right)$, yielding 
\begin{equation}
  \tau_{\rm E} \simeq \frac{\ln N}{\ln(\kappa/2)},
\end{equation}
up to an $N$-independent constant. Once the scattering matrix is
known, the conductance $G$ follows from the Landauer formula,
$G(\varepsilon)
= \sum_{m=1}^{N} \sum_{n=N+1}^{2N} |S_{mn}(\varepsilon)|^2$.

We study
transport properties as a function of time, simply by truncating the
sum in Eq. (\ref{eq:Smat}) after a time $t = t_0$. 
We calculate the ensemble averaged conductance by averaging over
$\varepsilon$, $\theta$, lead positions, and a small range of 
$\kappa$. An advantage of the time-resolved approach is the ability 
to evaluate the quasienergy average analytically. In order to 
achieve sufficient accuracy, we take between
$20\, 000$ and $80\, 000$ realizations
for $\theta$, the lead positions, and $\kappa$. 

Ideally, for two contacts with $N$ channels, the ensemble average of
the classical conductance is $N/2$ and WL can be calculated by taking
the difference between the quantum and classical conductances. In
practice, however, the ensemble average of the classical conductance
may differ from $N/2$ and this simple procedure fails.
Alternatively, one compares conductances with and without a
time-reversal symmetry breaking perturbation. 
However, such
a perturbation may affect the classical dynamics, complicating
the interpretation of the results. Instead, we choose to subtract the 
classical
conductance by comparing conductances at $N$ and $2N$ channels, for
the {\em same} realization of the classical parameters. The classical
conductance is proportional to the channel number, whereas the WL 
correction is independent of it.

\begin{figure}
\epsfxsize=0.95\hsize
\epsffile{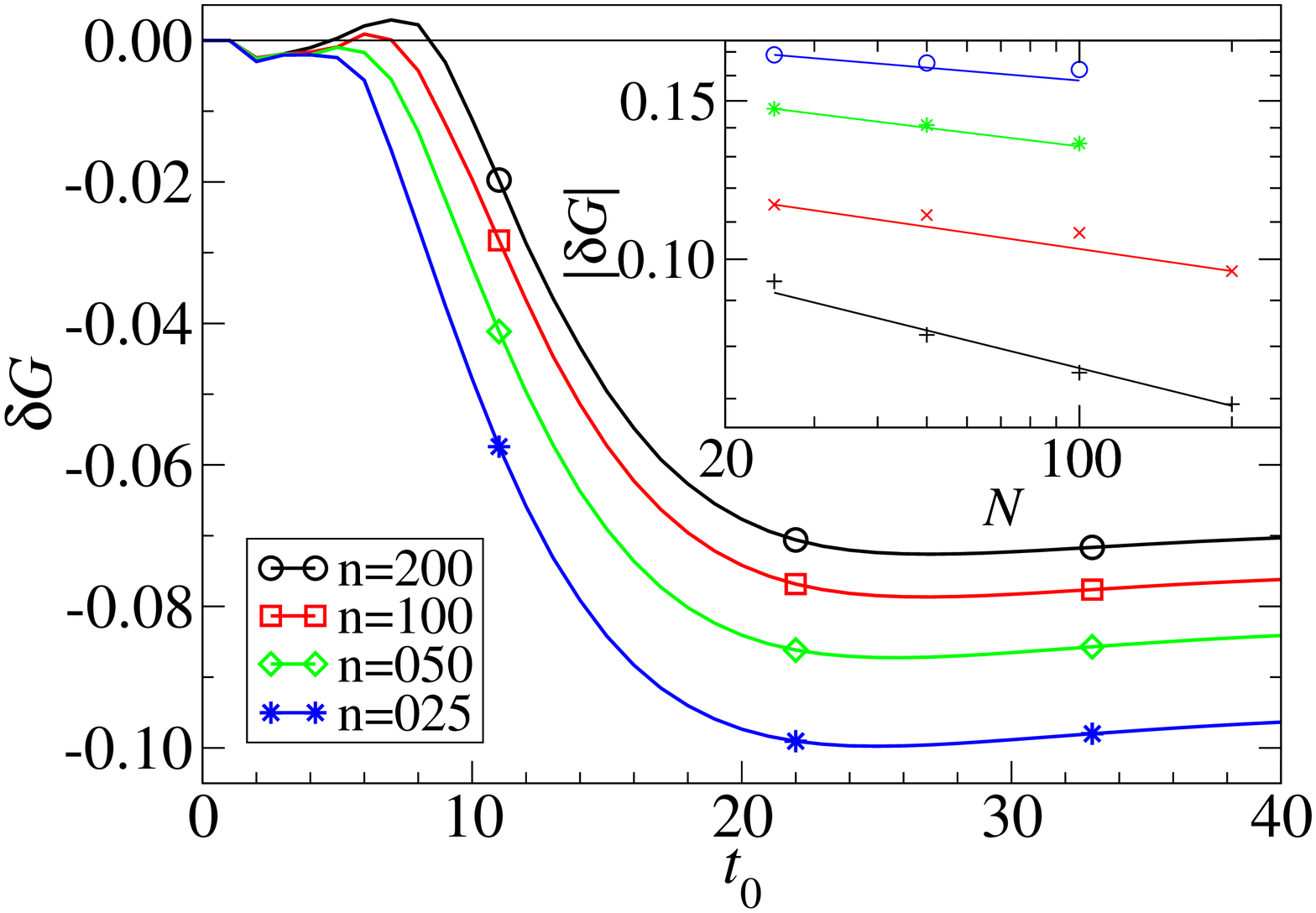}
\caption{\label{fig:wl1}
Time-resolved weak localization correction $\delta G$ as
a function of the cut-off time $t_0$, for $\kappa \approx 10$ and
$\tau_{\rm D} = 5$, and for four values of $N$. Inset: summary
of results for the weak localization correction. Data points are
obtained from the limit $t_0 \gg \tau_{\rm D}$ of simulations for
$\tau_{\rm D} = 5$ and $\kappa \approx 10$ ($+$), $\tau_{\rm D} = 5$
and $\kappa \approx 20$ ($\times$), $\tau_{\rm D} = 10$
and $\kappa \approx 10$ ($*$), and $\tau_{\rm D} = 10$ and
$\kappa \approx 20$ (o). The solid lines are fits $\propto
\exp(-\tau_{\rm E}/\tau_{\rm D})$, 
with $d\tau_{\rm E}/d\ln N$ taken 
from the low-$t_0$ part of the simulation data.}
\end{figure}

Simulation results for the WL correction to the 
conductance for $\tau_{\rm D} = 5$ and $\kappa \approx 10$
are shown in Fig.\ \ref{fig:wl1} as a function of the
cut-off time $t_0$. We also performed simulations for
$\tau_{\rm D} = 5$ and $\kappa \approx 20$, $\tau_{\rm D}
= 10$ and $\kappa \approx 10$, and $\tau_{\rm D}
= 10$ and $\kappa \approx 20$. In all cases, the 
derivative $d\tau_{\rm E}/d\ln N$, as extracted from the 
low-$t_0$ part of the curves, is consistent with an onset of
the WL correction after a time $2 \tau_{\rm E}$, in agreement
with the predictions of the semiclassical theory. A summary of our 
results for $\delta G$ is shown in the inset of Fig.\ \ref{fig:wl1}, 
together 
with a fit $\propto \exp(-\tau_{\rm E}/\tau_{\rm D}$), 
with $d\tau_{\rm E}/d\ln N$ 
obtained from the low-$t_0$ part of the data. 
The simulated
$\delta G$ has a systematic dependence on $\tau_{\rm E}$, 
consistent with the semiclassical theory. 
Previous numerical calculations of $\delta G$ did not show a 
dependence on $\tau_{\rm E}$ \cite{kn:tworzydlo2004b}, probably 
due to insufficient accuracy. The fact that
the over-all magnitude of WL is smaller than in Eq.\ (\ref{eq:dtresult}) 
is well known \cite{kn:tworzydlo2004b} and caused by a 
lack of ergodicity for the dwell times we consider.
Note that
our results for both the onset of WL and its magnitude
disagree with the `effective random matrix theory'.
We have further verified our results by calculating $\delta G$ 
for the three-kick model, a version of the open
quantum kicked rotator in which time-reversal symmetry can
be broken \cite{kn:tworzydlo2004b}. We found $\delta G = 0$ 
when the symmetry is broken. 

We believe that our simulation results, together with the corrected
semiclassical theory, resolves the controversy around the
Ehrenfest-time dependence of the weak localization correction to
the conductance. Our results quantitatively
confirm the corrected semiclassical
theory and disagree with the `effective random matrix theory'. 
The
results presented here are limited to weak localization and do
not carry over to conductance fluctuations. We have
also carried out numerical simulations of the variance of the 
conductance for the open quantum kicked rotator
\cite{kn:rahav2006}, and find behavior
consistent with that reported previously in the literature
\cite{kn:tworzydlo2004,kn:tworzydlo2004c,kn:jacquod2004}: no
dependence of conductance fluctuations on the Ehrenfest
time $\tau_{\rm E}$ and an 
onset time $\tau_{\rm E}$, not $2 \tau_{\rm E}$. Currently,
there is no semiclassical theory these results can be compared
with.

We would like to thank C.\ Beenakker, S.\ Fishman, H.\ Schomerus,
P.\ Silvestrov, and D.\ Ullmo for discussions. 
This work was supported by the NSF
under grant no.\ DMR 0334499 and by the Packard Foundation.

%\bibliography{/afs/msc.cornell.edu/home/brouwer/brouwer/grant/2003/refs}

\end{document}